\documentclass{ws-ijmpa}
\usepackage{lineno}
\usepackage{color}

\newcommand{\piz}{\pi^0}
\newcommand{\etap}{\eta^{\prime}}
\newcommand{\EE}{e^+e^-}
\newcommand{\pp}{\pi^+\pi^-}
\newcommand{\beq}{\begin{equation}}
\newcommand{\eeq}{\end{equation}}
\newcommand{\beqn}{\begin{eqnarray}}
\newcommand{\eeqn}{\end{eqnarray}}
\newcommand{\beqns}{\begin{eqnarray*}}
\newcommand{\eeqns}{\end{eqnarray*}}
\newcommand{\bfg}{\begin{figure}}
\newcommand{\efg}{\end{figure}}
\newcommand{\bitm}{\begin{itemize}}
\newcommand{\eitm}{\end{itemize}}
\newcommand{\bnum}{\begin{enumerate}}
\newcommand{\enum}{\end{enumerate}}
\newcommand{\btbl}{\begin{table}}
\newcommand{\etbl}{\end{table}}
\newcommand{\btbu}{\begin{tabular}}
\newcommand{\etbu}{\end{tabular}}
\def\braket#1{|#1\rangle}

\def\eref#1{(\ref{#1})}

\begin{document}
\linenumbers

\title{A mathematical review on the multiple-solution problem\footnote{Electronic version
of this paper has been published as [International Journal of Modern Physics A (IJMPA)
, Volume No.26, Issue No. 25., Year 2011, Page
  4511-4520] [Article DOI 10.1142/S0217751X11054589] ©
[copyright World Scientific Publishing Company] [http://www.worldscinet.com/ijmpa]} }

\author{K. Zhu\footnote{e-mail: zhuk@ihep.ac.cn}, X.~H. Mo, C.~Z. Yuan,
P. Wang}
\address{Institute of High Energy Physics, Chinese Academy of
Sciences,\\ P.O. Box 918(1), Beijing 100049, China}
\maketitle


\begin{abstract}

The recent multiple-solution problem in extracting physics
information from a fit to the experimental data in high energy
physics is reviewed in a mathematical viewpoint. All these
multiple solutions were found via a fit process previously, while
in this letter we prove that if the sum of two coherent
Breit-Wigner functions is used to fit the measured distribution,
there should be two and only two non-trivial solutions, and they
are related to each other by analytical formulae. For real
experimental measurements in more complicated situations, we also
provide a numerical method to derive the other solution from the
already obtained one. The excellent consistency between the exact
solution obtained this way and the fit process justifies the
method. From our results it is clear that the physics
interpretation should be very different depending on which
solution is selected. So we suggest that all the experimental
measurements with potential multiple solutions be re-analyzed to
find the other solution because the result is not complete if only
one solution is reported.

\end{abstract}

\ccode{PACS numbers: 02.30.Fn, 02.60.Cb, 13.66.Bc, 14.40.-n}

\section{Introduction}

Interference as a nature phenomenon has been observed for a very
long time in situations where waves intersect, no matter the
mediate material is water, string, sound, or light. It has been
studied in depth and also widely used in a range of physical and
engineering measurements. However, classic physics and quantum
mechanics provide basically different explanations of this
phenomenon. In classic physics, if two meeting waves are
considered contributing to a process, what observed is just the
sum of amplitudes of two waves, i.e., $A(x) = A_1(x) + A_2(x)$,
where $x$ is a generalized coordinate which could be position,
momentum, time, energy, etc. But in quantum mechanics, their wave
functions are summed to obtain the total amplitude (generally
there are relative phases between them), i.e. $| \psi(x)\rangle =
| a(x) \rangle + | b(x) \rangle$. And the experimentally measured
quantities are usually proportional to the modulus of the
amplitude squared, and thus one generally has contribution from an
interference term $\langle a | b \rangle$. Compared with the
classic ones, many new and fantastic features are caused by this
additional interference term $\langle a | b \rangle$, and the
ambiguity in extracting information from observation is one of
them.

Usually the experimental quantities depending on $|\psi|^2$ are
measured, and from which we extract the information of the
amplitudes $\braket{a}$ and $\braket{b}$. Unlike in the classic
physics case, as there is a square operation between the
observable and the amplitudes, we would expect other solutions,
$\braket{a'}$ and $\braket{b'}$, be found in extracting amplitudes
from physics measurements. It is true that the existing freedom on
the global phase is non-relevant to the physics in this extraction
procedure. However, more and more experimental analyses presented
recently imply that different solutions with different relative
phases would lead to non-trivial different physics
interpretations.

Some earlier examples reporting multiple solutions are in the
study of the so-called $Y$ states via initial state radiation
($ISR$) by the Belle experiment\cite{:2007sj,:2007ea}. The
invariant mass distributions of $\pi^+\pi^-J/\psi$ and
$\pi^+\pi^-\psi(2S)$ are fitted with two coherent resonant terms
and an incoherent background term. Another example is the study of
the decay dynamics of $\etap\to \gamma\pp$ mode. When the $\pp$
invariant mass distribution is fitted with a coherent sum of the
$\rho$ resonance and a contact term, two solutions are found with
one solution corresponding to constructive interference between
the two amplitudes while the other corresponding to destructive
interference.\cite{Chen:2007nb} Some recent examples are presented
in Refs.~\refcite{Yuan:2009gd} and \refcite{Mo:2010bw}. In
Ref.~\refcite{Yuan:2009gd} two solutions are found for both the
branching fraction measurement of $\phi \to \omega \pi^0$ and the
$\rho-\omega$ mixing study. In Ref.~\refcite{Mo:2010bw}, four sets
of solutions are found in fitting the $R$-values to extract the
resonant parameters of the excited $\psi$ states, namely the
$\psi(4040)$, $\psi(4160)$, and $\psi(4415)$.

However, it is notable that in
Refs.~\refcite{:2007sj}--\refcite{Shen:2009mr} all the multiple
solutions are found via fitting process. And we know fit method
always suffers from background uncertainties and limited
statistics. Then some interesting questions are raised naturally
such as: whether they are exact solutions or only approximate
results due to the shortcomings of the fit process or statistical
fluctuation; whether these solutions always exist or just appear
in some special cases; how many fold ambiguities there are if
multiple-solution exists; and if one special solution has already
been found, whether the others can be derived from it. Some of
these questions are explored from physics point of view in
Refs.~\refcite{Yuan:2009gd} and \refcite{Mo:2010bw}, and
mathematical attempts are described in Ref.~\refcite{Bukin:2007kx}
and \refcite{Yuan:2010sk}. Following the clue of these above
mentioned studies we perform the present investigations in this
paper. Comparing with Ref.~\refcite{Bukin:2007kx} in which
Fourier transformation is applied, the solution finding process is
extremely simplified in this study; comparing with
Ref.~\refcite{Yuan:2010sk}, more general conclusions are obtained
in this analysis.

In section~\ref{secII}, at the outset, a general and mathematic
model for the sum of two amplitudes is established on the basis of
the known facts from physical analyses. If two amplitudes are both
the commonly used Breit-Wigner (BW) functions, the analytical
expression for the two solutions are obtained. Moreover, an
effective approach is developed for deriving the algebra equations
related to the solutions. Then many double-solutions are deduced
for distinctive forms of amplitude functions. After that, we put
forth a constraint on the ratio of the two amplitude functions,
which ensures that there will be non-trivial double solutions. In
section~\ref{secIII}, we use a toy numerical example to check and
confirm our results. We also develop a numerical procedure to get
the unknown solution from the known one when the form of the
amplitude function is extremely complicated. Finally there is a
short discussion on the consequence of having multiple solutions
in extracting physics information from the experimental data, and
we also present our suggestion on how to handle this situation.

\section{Mathematical methodology}
\label{secII}

If scrutinizing the relevant analyses with multiple
solutions~\cite{:2007sj}\cdash\cite{Shen:2009mr}, we can note two
prominent characteristics: 1.  all set of solutions have equal
goodness-of-fit; 2. although all parameters including the masses,
the total widths, the partial widths and some other related
parameters are allowed to float in the fit, it is observed that
the differences between multiple solutions are only in the partial
widths and the relative phases between the amplitudes. The first
point indicates that all solutions are mathematically equivalent
while the second point implies that the main difference for
different solutions is in the normalization factor and the
relative phase between them. In the light of these experimental
facts, we abstract a general mathematical model for multi-solution
problem. Without losing generality, the study that follows focuses
on the case of two amplitude functions.

\subsection{Solutions for two BW amplitudes}

Generally, a sum of two quantum amplitudes can be described by a
complex function $e(x,z_1,z_2)$ with form
\begin{linenomath*}
\begin{equation}
\label{eq:gen}
 e(x,z_1,z_2) = z_1\ g(x) + z_2\ f(x) ~,
\end{equation}
\end{linenomath*}
where $g(x)$ and $f(x)$ are both complex functions, $x$ is a real
variable, and $z_1$, $z_2$ are complex numbers. Our goal is to
find non-trivial different series of parameters $z'_1$ and $z'_2$
that satisfy
\begin{linenomath*}
\begin{equation}
 \left| e(x,z_1,z_2) \right|^2 = \left| e(x,z'_1,z'_2) \right|^2~.
\label{eq:sol}
\end{equation}
\end{linenomath*}
Noticed that the global phase plays no role in the amplitude
squared, we can reduce the dimension of $z_1-z_2$ parameter space
to a $z-d$ space in which $d$ is real number, and re-write $\left|
e(x,z_1,z_2) \right|^2$ to a more convenient form by defining
\begin{linenomath*}
\begin{eqnarray}
&&\left| e(x,z_1,z_2) \right|^2 \equiv \frac{1}{d}\left|
g(x) + z\ f(x) \right|^2 = \frac{\left| g(x) \right|^2}{d} \left| 1+z
\frac{f(x)}{g(x)} \right|^2 \cr
&&\equiv \frac{\left| g(x) \right|^2}{d} \left| 1+z\ F(x) \right|^2
    \equiv  \frac{\left| g(x) \right|^2}{d} E(x,z) ~.
\label{eq:eandeprm}
\end{eqnarray}
\end{linenomath*}
Here $F(x) \equiv f(x)/g(x)$ and $E(x,z) \equiv \left| 1+ z\
F(x)\right|^2$. Since $\left| g(x)\right|^2$ is only a multiply
factor and is independent of $d$ and $z$, it can be dropped in the
following discussion. Now we only focus on finding different
series of $d$ and $z$ that keep $E(x,z)/d$ unchanged. Denoting the
real and imaginary parts of $F(x)$ with $R_F(x)$ and $I_F(x)$, as
well as $R_z$ and $I_z$ for $z$, respectively, and expressing
$E(x,z)$ by these real and imaginary components, we obtain
\begin{linenomath*}
\begin{equation}
  E(x,z) = (R_F^2 + I_F^2) (R_z^2 + I_z^2) - 2 I_F I_z + 2 R_F R_z + 1~.
\label{eq:main}
\end{equation}
\end{linenomath*}
For compactness, the explicit dependence on $x$ of $R_F(x)$ and
$I_F(x)$ is removed here. Without losing generality, set $d=1$ as
an initial solution for convenience, so our task is to find all
possible $d'$ and $z'$ to render $E(x,z')/d' = E(x,z)$. To
specialize our work, we consider the case when both $g(x)$ and
$f(x)$ are non-relativistic BW functions\cite{Perkins:1982xb}
\begin{linenomath*}
 $$ g(x) = \frac{\Gamma_{g}}{(x-M_{g})+i\Gamma_{g}}~, $$
 $$ f(x) = \frac{\Gamma_{f}}{(x-M_{f})+i\Gamma_{f}}~, $$
\end{linenomath*}
where $M$ and $\Gamma$ are the mass and width of a resonance,
respectively. This BW-form amplitude function is chosen because
it's broadly adopted in high energy physics. With the above forms
of $g(x)$ and $f(x)$, the real and imaginary components of $F(x)$
are
\begin{linenomath*}
 $$
 R_F =\frac{\Gamma_f[ \Gamma_g \Gamma_f + (M_g -x)(M_f -x)]}
 {\Gamma_g [\Gamma_f^2 + (M_f-x)^2]} ,
 $$
\end{linenomath*}
and
\begin{linenomath*}
$$
 I_F =\frac{\Gamma_f[ \Gamma_f (M_g -x) - \Gamma_g (M_f -x)]}
 {\Gamma_g [\Gamma_f^2 + (M_f-x)^2]} ~.
 $$
\end{linenomath*}
After some algebra, we get an interesting relation
\begin{linenomath*}
\begin{equation}
R_F^2 + I_F^2 = a R_F + b I_F + c \;,
\label{eq:abc}
\end{equation}
\end{linenomath*}
with
\begin{linenomath*}
 \begin{equation}
 a = \frac{\Gamma_g + \Gamma_f}{\Gamma_g} ,\;\;
 b = \frac{M_g - M_f}{\Gamma_g} ,\;\;
 c = -\frac{\Gamma_f}{\Gamma_g} \;.
 \end{equation}
\end{linenomath*}
With Eq.~\eref{eq:abc}, $E(x,z)$ is recast as
\begin{linenomath*}
\begin{equation}
\label{eq:dbw}
 R_F(aR_z^2 + aI_z^2 + 2R_z) + I_F(bR_z^2 + bI_z^2 - 2I_z ) +
 c(R_z^2 +I_z^2)+1 .
\end{equation}
\end{linenomath*}
Similar expression can be obtained for $E(x,z')$. Notice that
$R_F$ and $I_F$ are functions in variable space ($x$ space), while
$R_z$ and $I_z$ are functions in parameter space ($z-d$ space), if
we want $E(x,z')/d' = E(x,z)$ for any $x$, then the corresponding
functions in parameter space (the coefficients of the functions in
variable space) should be equal. This requirement immediately
yields
\begin{linenomath*}
 \beq
\begin{array}{rcl}
 aR_{z'}^2 + aI_{z'}^2 + 2R_{z'} &=& d'(aR_z^2 + aI_z^2 + 2R_z)~, \\
 bR_{z'}^2 + bI_{z'}^2 - 2I_{z'} &=& d'(bR_z^2 + bI_z^2 - 2I_z)~, \\
 cR_{z'}^2 + cI_{z'}^2 + 1       &=& d'(cR_z^2 + cI_z^2 + 1 ) .
\end{array}
\label{eq:bwary}
 \eeq
\end{linenomath*}
In the light of the set of equations, it turns out that $d'$
must satisfy a second order equation and there are two roots of
it. One is the trivial solution with $d'=1$ and $z'=z$
correspondingly, and the other one is
\begin{linenomath*}
\begin{eqnarray}
 d' &=&\frac{a^2+b^2+4c}{(a-2R_zc)^2 + (b+2I_zc)^2} \cr
 z' &=&\left(R_z d' - \frac{a(d'-1)}{2c}\right) +
      \left(I_z d' + \frac{b(d'-1)}{2c}\right)i
\label{eq:dz}
\end{eqnarray}
\end{linenomath*}

\subsection{Some special solutions}

In this section we consider some other forms for $f(x)$ and
$g(x)$. The first form is from the KLOE experiment on $e^+ e^- \to
\omega \pi^0$ with both $\omega \to \pi^+ \pi^- \pi^0$ and $\omega
\to \gamma \pi^0$. The cross section of $\EE\to \omega\piz$ in the
vicinity of the $\phi$ resonance, as a function of the
center-of-mass energy, $\sqrt{x}$, is parameterized as
\begin{linenomath*}
 \beq
 \label{xsection}
 \sigma(\sqrt{x}) = \sigma_{nr}(\sqrt{x})
 \cdot\left|1-z\frac{M_{\phi}\Gamma_{\phi}}{D_{\phi}(\sqrt{x})}\right|^2
 \eeq
\end{linenomath*}
in Ref.~\refcite{Ambrosino:2008gb}, where $\sigma_{nr}(\sqrt{x}) =
\sigma_{0} + \sigma' (\sqrt{x} - M_\phi)$ is the bare cross
section for the non-resonant process, parameterized as a linear
function of $\sqrt{x}$; $M_{\phi}$, $\Gamma_{\phi}$, and
$D_{\phi}=M_{\phi}^2-x-iM_{\phi}\Gamma_{\phi}$ are the mass, the
width, and the inverse propagator of the $\phi$ meson,
respectively. Here $z$ is a complex number which depicts the
interference effect. Comparing with definition of $E(x,z)$, $f(x)$
and $g(x)$ have the forms
\begin{linenomath*}
 $$ g(x)=-1~,
 f(x)= \frac{M_{\phi}\Gamma_{\phi}}{M_{\phi}^2-x-
 iM_{\phi}\Gamma_{\phi}}~.
 $$
\end{linenomath*}
The simple algebra yields $R_F^2 + I_F^2 = -I_F$, which in turn
gives
\begin{linenomath*}
 \beq
 \begin{array}{rcl}
  R_{z'}^2 + I_{z'}^2 + 2I_{z'} &=& d' (R_z^2 + I_z^2 + 2I_z)~, \\
  R_{z'} &=& d' R_z~,  \\
  1 &=& d'~.
 \end{array}
 \label{eq:kloeary}
 \eeq
\end{linenomath*}
With the last equality $d'=1$, the relation $R_{z'} = d' R_z$
implies $R_{z'} = R_z$, then the first equation provides the other
non-trivial solution
\begin{linenomath*}
$$ z'=R_z - i (I_z +2 )~. $$
\end{linenomath*}
They are just the results acquired in Ref.~\refcite{Yuan:2010sk}
by another method.

As the second example, we consider the form
\begin{linenomath*}
 \beq
 g(x)=\frac{1}{x}~,~~
 f(x)= \frac{1}{m^2-x+im\Gamma}~,
 \label{xct_omg}
 \eeq
\end{linenomath*}
which is usually used to extract the resonant information of
$\omega$ in fitting the data of $\EE \to \pp$ cross
section\cite{Akhmetshin:2006bx}. Here $m$ and $\Gamma$ indicate
the mass and total decay width of the resonance. Accordingly, we
obtain $R_F^2 + I_F^2 = -R_F+ \zeta I_F (\zeta=-m/\Gamma)$, which
in turn gives
\begin{linenomath*}
 \beq
\begin{array}{rcl}
 R_{z'}^2 + I_{z'}^2 - 2R_{z'} &=& d' (R_z^2 + I_z^2 - 2R_z)~, \\
 \zeta (R_{z'}^2 + I_{z'}^2) - 2I_{z'} &=& d' [ \zeta( R_z^2 + I_z^2) - 2I_z]~, \\
 1 &=& d'~.
\end{array}
\label{eq:kloearyII}
 \eeq
\end{linenomath*}
After some algebra, we get the other non-trivial solution
\begin{linenomath*}
\begin{equation}
z'=\frac{2+(\zeta^2-1)R_z-2\zeta I_z}{1+\zeta^2} + i {
  \frac{2\zeta (1-R_z)-(\zeta^2-1)I_z}{1+\zeta^2}} ~.
\label{eq:sol-of-cloeii}
\end{equation}
\end{linenomath*}

We consider a more general case, when $f(x)$ and $g(x)$ are any
non-trivial functions, but their ratio must ensure that the real
or imaginary component of $F(x)$ is constant\footnote{This can be
realized, for example, if $f(x)=\rho_f(x) e^{i\theta_f(x)}$ and
$g(x)=\rho_g(x) e^{i\theta_g(x)}$, where $\rho_{f,g}(x)$ and
$\theta_{f,g}(x)$ are any non-trivial functions. As long as there
exist relations $\rho_{f}(x) = \rho_{g}(x)\cdot
\sqrt{h^2(x)+\kappa^2}$ and $\theta_{f}(x) = \theta_{g}(x) +
\tan^{-1}(h(x)/\kappa)+2\pi n $ ($n$: any integer), it always has
$F(x)= \kappa + i h(x)$.}. In this case, there exist two
solutions. Specially speaking, when $F(x)= \kappa + i h(x)$, with
$h(x)$ being a non-trivial function and $\kappa$ is a non-zero
real constant. Here
\begin{equation}
E(x,z)=h^2(x)(R^2_z+I^2_z) - 2h(x)I_z + \kappa(R^2_z +I^2_z) + 2\kappa
R_z +1 ~.
\end{equation}
From $E(x,z')/d'=E(x,z)$ and we require all coefficients of each
order of $h(x)$ being equated respectively, i.e.
\begin{eqnarray}
R^2_{z'}+I^2_{z'} &=& d'(R^2_z + I^2_z) ~, \cr
-2I_{z'} &=& d'(-2I_z) ~,\cr
\kappa^2(R^2_{z'}+I^2_{z'}) + 2\kappa R_{z'}+1 &=&
d'[\kappa(R^2_z+I^2_z) + 2\kappa R_z + 1] ~.
\end{eqnarray}
Besides the trivial solution $d'=1$ and $z'=z$, there exists the
other solution to above set of equations,
\begin{linenomath*}
\begin{eqnarray}
 d'&=&\frac{1}{4 \kappa^2(R_z^2+I_z^2)+4 \kappa R_z + 1} ~, \cr
 z' &=& -d'[2 \kappa(R^2_z+I^2_z)+R_z] + i(I_z d') ~.
 \label{eq:groff}
\end{eqnarray}
\end{linenomath*}
When $F(x)= h(x) + i\kappa$, via similar derivation,
the other non-trivial solution is obtained as
\begin{linenomath*}
\begin{eqnarray}
 d' &=&\frac{1}{4 \kappa^2(R_z^2+I_z^2)-4 \kappa I_z + 1} ~, \cr
 z' &=& R_z d' + id'[ 2\kappa(R^2_z+I^2_z)- I_z] ~.
 \label{eq:gioff}
\end{eqnarray}
\end{linenomath*}

There are also other cases where there are two solutions, such as
when $R_F$ is a linear function of $I_F$, or vice versa, we will
not discuss them in details here.

\subsection{Constraint on the amplitude functions}

Despite of the examples shown in the previous section, it is clear
that the double-solution issue is not universal for any forms of
the functions. Actually, it is easy to find some forms of $f(x)$
and $g(x)$, in which no multiple-solutions can be found, $g(x)=x$
and $f(x)=x^3+ix^2$ is such an example.

This is why, although still far from the final answer, we want to
discuss what kind of constraints can be applied to the functions
if double-solutions exists, i.e. what kind of amplitude functions,
$f(x)$ and $g(x)$ in the preceding section, will guarantee double
solutions. When we return to the study of two BW amplitudes, we
notice that the relation in Eq.~\eref{eq:abc} is crucial for
obtaining the double-solutions, and this relation provides a
constraint on $F(x)$. It is easy to check that all the special
forms with double-solutions found by us, obey this requirement.
In short: there will be two solutions if the Argand diagram of
$F(x)$ is a circle.

\section{Check and Application}
\label{secIII}

As a cross check, let's consider an ad hoc example: the parameters
of the two BW functions and one solution are set as
\begin{linenomath*}
 $$
 M_g = 3.0,\;\; \Gamma_g = 0.4 ,\;\; M_f=2.1,\;\; \Gamma_f=0.1, \;\;
 z=1-i\;.
 $$
\end{linenomath*}
Using the aforementioned method, we can find another solution,
which is exactly repeated by fitting with maximum likelihood
method. The comparison of the results is shown in
Table~\ref{tab:example1}.

\begin{table}[htb]
\tbl{Comparison between exact solution and that obtained from fit
process in the case of summing up two simple BW functions. For the
fit, toy MC is used to generate a 10,000 events data sample. }
{\begin{tabular}{@{}ccccc@{}}
 \toprule
 Item       & Input  & The other sol. & Fit I & Fit II \\ \colrule
$d$        &  $1$   & $0.529$ & $-$ &  $-$  \\
$R_z$      &  $1$   & $0.647$ & $1.019\pm0.054$  &  $0.644\pm0.040$ \\
$I_z$      &  $-1$  & $1.588$ & $-1.019\pm0.060$ &  $1.601\pm0.028$ \\ \colrule
$M_g$      &  $3.0$ & $3.0$   & $3.011\pm0.010$ & $3.011\pm0.010$  \\
$\Gamma_g$ &  $0.4$ & $0.4$   & $0.402\pm0.010$ & $0.402\pm0.010$  \\
$M_f$      &  $2.1$ & $2.1$   & $2.101\pm0.003$ & $2.101\pm0.003$  \\
$\Gamma_f$ &  $0.1$ & $0.1$   & $0.101\pm0.003$ & $0.101\pm0.003$  \\
\botrule
\end{tabular}
\label{tab:example1}}
\end{table}

This example indicates that in principle, the fit procedure can be
used as a feasible approach to find the multiple solutions from
the experimental data.

It is obvious that, for the two BW amplitudes case, if one
solution is obtained by fitting to the data, the other one can be
readily and analytically obtained by applying Eq.~\eref{eq:dz}.
This definitely saves a lot of time and energy.  However, due to
the complexity of the expressions in practice, the solution has to
be obtained from the following numerical method. 
Firstly, we draw $F(x)$ in the complex plane to check whether
it is a circle. If the answer is yes, we need to determine the
parameters $a$, $b$, and $c$ in relation $R_F^2(x)+I_F^2(x) = a
R_F(x) + b I_F(x) +c $. This can be achieved by randomly selecting
three points on the curve, i.e. set $x$ in $F(x)$ with three
randomly chosen values, to obtain three linear equations with
$a$, $b$, and $c$ as variables, whose values can be got by solving
the set of equations. This process should be realized
numerically when $F(x)$ takes a very complicated form in practice.
With $a$, $b$, and $c$ obtained, we can derive the other solution
with Eq.~\eref{eq:dz} as was shown before.  We illustrate this
method by the examples which are selected from the initial state
radiation measurements at BaBar and
Belle\cite{:2007sj,Liu:2008hja}, where the $\pi^+ \pi^- \psi(2S)$
and $\pi^+ \pi^- J/\psi$ invariant mass distributions are
described by two coherent resonances. The cross sections are
formulated as
\begin{linenomath*}
 $$
  \sigma(s) = |BW_1(s)+BW_2(s)\cdot e^{i\phi}|^2 \;,
 $$
\end{linenomath*}
where $BW_1$ and $BW_2$ represent the two resonances and $\phi$ is
the relative phase between them. Then in our
frame $BW_1$, $BW_2$, and $e^{i\phi}$ represents $g(x)$, $f(x)$,
and $z$ respectively, and $F(x)=BW_2/BW_1$.  And the BW form of a
single resonance in these two papers is
\begin{linenomath*}
 \begin{equation}
 BW(s)=\sqrt{\frac{M^2}{s}} \frac{\sqrt{12\pi \Gamma_{e^+e^-}
 B(R\to f)\Gamma_{tot}}}
 {s-M^2+i M\Gamma_{tot}} \sqrt{\frac{PS(s)}{PS(M)}} ~,
 \end{equation}
\end{linenomath*}
where $M$ is the mass of the resonance, $\Gamma_{tot}$ and
$\Gamma_{e^+e^-}$ are the total width and partial width to
$e^+e^-$, respectively, $B(R\to f)$ is the branching fraction of
$R$ decays into final state $f$, and $PS(s)$ is the three-body
decay phase space factor.  Noticed here that
$F(x)$ is very complicated. We draw it in the complex plane and
find it is in a good circle shape, then we can obtain the
parameters $a$, $b$, and $c$ from this plot. Using
Eq.~(\ref{eq:dz}) and the first solution as input we obtain the
second solution as shown above; or reversely, using the second
solution as input to obtain the first one. All the results from
our method and comparisons with the experimental fits are shown in
Table~\ref{tab:comp}, where only the statistical errors are quoted
for experimental measurements and only the central values are used
as input to get the other solution. From the Table it is clear
that our results reproduce the results from the fit process very
well, and we consider this as a justification of our method. 
\begin{table}[htb]
\tbl{Comparison of the exact solutions with
those obtained from the fit process for two real experimental
measurements. For experimental results, only statistical errors
are quoted; for our method, solution I is obtained with solution
II as input, and vice versa. }
{\begin{tabular}{@{}ccccccc@{}} \toprule
 Solutions & I &      &       &    II &   &   \\ \colrule
 Items & $B\Gamma_{e^+e^-}(R1)$  & $B\Gamma_{e^+e^-}(R2)$& $\phi$ &
 $B\Gamma_{e^+e^-}(R1)$  & $B\Gamma_{e^+e^-}(R2)$& $\phi$ \\ \colrule
 fit results in Ref.~\refcite{:2007sj} &
 $5.0\pm1.4$ & $6.0\pm1.2$ & $12\pm29$ &
 $12.4\pm2.4$ &  $20.6\pm2.3$ & $-111\pm7$ \\
 by our method
 & $4.7$ &  $6.4$ & $12$  & $12.8$ &  $20.4$ & $-111$ \\ \colrule
 fit results in Ref.~\refcite{Liu:2008hja}  &
 $11.1^{+1.3}_{-1.2}$ &  $2.2^{+0.7}_{-0.6}$ & $18^{+23}_{-24}$ &
 $12.3\pm1.2$ &  $5.9\pm1.6$ & $-74^{+16}_{-12}$ \\
 by our method  & $11.1$ &  $2.1$ & $18$
 & $12.3$ &  $6.0$ & $-74$ \\
\botrule
\end{tabular} \label{tab:comp}}
\end{table}

\section{Discussion}

As been found, when the measured distribution is described by $|g(x)+z
f(x)|^2/d$ and $F(x)=f(x)/g(x)$ fulfills the relation of
Eq.~\eref{eq:abc}, i.e., $F(x)$ is a circle in complex plane, there
are and only are two non-trivial solutions. It has also been proved
that if $f(x)$ and $g(x)$ are both simple BW functions, this relation
is exactly satisfied and Eq.~\eref{eq:dz} can be utilized to derive
the other solution analytically from the one obtained from the
fit. For other transmogrified BW functions, which could be extremely
complicated, the relation Eq.~\eref{eq:abc} still hold for $F(x)$
by numerical checking. So there will be double solutions for these
forms too and with $a$, $b$, and $c$ obtained numerically the other
solution can be derived by using the same method. The excellent
consistency between our exact solutions and experimental fit results
justifies this method. We should mention that when we use the
experimental fit results as input, only the central values are
considered to prove the method is valid. In principle, one can obtain
the errors of the parameters in the second solution given the full
covariant matrix is available for the first solution (unfortunately,
very often, experimental papers only give diagonal errors, and the
correlations between the variables are not reported). The uncertainty
of the numerical method is not discussed in case of complicated BW
forms, as the main purpose of this paper is to examine whether the
other solution exist or not, and how to find it.

We also notice that for both solutions, the parameters of each
resonance are same but the normalization factors. This implies that
the couplings to decay channels are different for different solutions
and some experimental reports may not be complete if only one solution
was reported while there are two-fold ambiguities. So we suggest any
experiment measurement with potential multiple-solution problem redo
the analysis to find out the other solutions. Finally, we should point
out that from Eq.~\eref{eq:main} we may find more conditions where
double solutions exist. For example, if the real or virtual component
of $F(x)$ is zero or the real component of $F(x)$ is a linear function
of the virtual one, there should be double solutions too.  However,
they are not normal in high energy physics so we do not discuss them
in detail here. Furthermore, only the sum of two coherent amplitudes
is considered in this paper, the generalization to more amplitudes is
still in progress.

\section*{Acknowledgments}

This work is supported in part by the National Natural Science
Foundation of China under Contracts Nos. 10775412, 10825524,
10935008, and 11005115, the Instrument Developing Project of the
Chinese Academy of Sciences under Contract No. YZ200713, Major
State Basic Research Development Program under Contracts Nos.
2009CB825203 and 2009CB825206, Knowledge Innovation Project of the
Chinese Academy of Sciences under Contract No. KJCX2-YW-N29 and
Innovation Project of Youth Foundation of Institute of High Energy
Physics under Contract No. H95461B0U2.

\end{document}